# Microscopic Metavehicles Powered and Steered by Embedded Optical Metasurfaces

Daniel Andrén[†,*], Denis G. Baranov[†], Steven Jones[†], Giovanni Volpe[‡], Ruggero Verre[†], Mikael Käll[†,*]

[†]Department of Physics, Chalmers University of Technology, S-412 96 Gothenburg, Sweden

[‡]Physics Department, Gothenburg University, S-412 96 Gothenburg, Sweden

*Corresponding authors: daniel.andren@chalmers.se, mikael.kall@chalmers.se

Nanostructured dielectric metasurfaces offer unprecedented opportunities to manipulate light by imprinting an arbitrary phase-gradient on an impinging wavefront.[1] This has resulted in the realization of a range of flat analogs to classical optical components like lenses, waveplates and axicons.[2–6] However, the change in linear and angular optical momentum[7] associated with phase manipulation also results in previously unexploited forces acting on the metasurface itself. Here, we show that these optomechanical effects can be utilized to construct optical metavehicles – microscopic particles that can travel long distances under low-power plane-wave illumination while being steered through the polarization of the incident light. We demonstrate movement in complex patterns, self-correcting motion, and an application as transport vehicles for microscopic cargo, including unicellular organisms. The abundance of possible optical metasurfaces attests to the prospect of developing a wide variety of metavehicles with specialized functional behavior.

One of the most distinguishing features of optical metasurfaces is their ability to simultaneously manipulate a wave's propagation direction and polarization despite being sub-wavelength in thickness. This results in an exchange of linear and angular momentum between light and matter and, by virtue of Newton's third law, optical forces and torques acting on the metasurface. Two main challenges need to be overcome in order to exploit this phenomenon for the practical realization of microparticles capable of movement and steering in a plane-wave light field (**Figure 1**). First, the microparticle should contain a metasurface that efficiently bends light towards a specific deflection angle $\theta$, such that a directed reaction force is induced from conservation of linear optical momentum. Second, the metasurface should be able to alter the angular momentum of light in such a way that a directed reaction torque $\tau_z$ appears; this will allow the microparticle to be steered through control of the incident polarization while being driven forward. We fulfill these criteria by incorporating a high-index dielectric metasurface (inset **Figure 1a**), constructed as a lattice of anisotropic metaatoms,[8] into a low-index transparent host particle, forming a *metavehicle* (inset **Figure 1b**). Directed propulsion is achieved by allowing only one of the principal lattice axes to support propagating first-order diffraction, and by designing each metaatom as an asymmetric dimer nanoantenna able to scatter preferentially into only one of these orders. The dimer nanoantennas are elongated along the propulsion direction, resulting in an anisotropic polarization response (shape birefringence) causing alignment of the metavehicle along the incident (linear) polarization vector[9,10] (see **Methods** section for a detailed theoretical treatment).

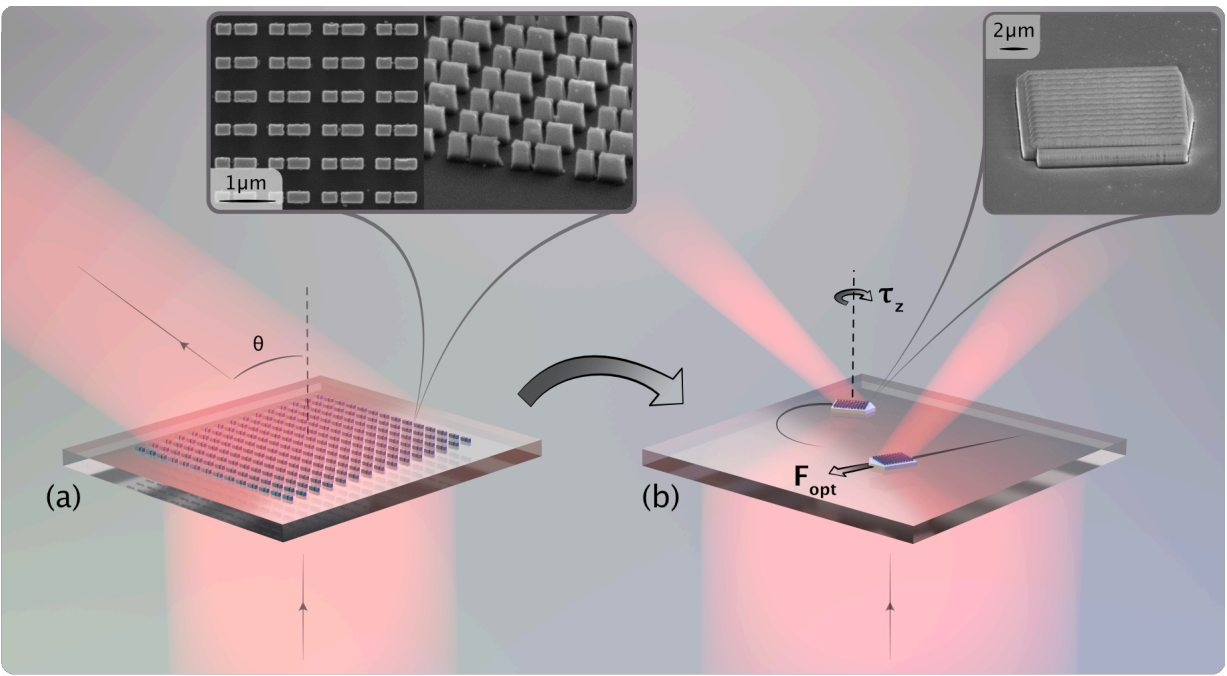

Figure 1: From stationary metasurfaces to mobile metavehicles. (a) Optical metasurfaces can manipulate the direction and polarization of light, which makes it possible to fabricate ultrathin optical components, but they are typically viewed as stationary objects that do not themselves react to the incident light. (b) However, by incorporating an engineered metasurface into a microparticle, the change in linear and angular momentum produced by the metasurface can be employed to propel and steer the microparticle. We call this a *metavehicle*. The insets display SEM images of a fabricated metasurface functioning as an anisotropic directional beam-deflector (left), and a metavehicle containing the same kind of metasurface (right).

The metavehicles were designed for operation in water using 1064 nm plane-wave illumination. The high-index metasurface consists of polycrystalline Si while the low-index host material is $SiO_2$. Fabrication was based on a combination of material deposition, electron-beam lithography and anisotropic etching, followed by suspension in water (**Methods** and **Supporting Figure S1**). The lattice constants are set due to fabrication constrains to $\Lambda_y$ = 600 nm and $\Lambda_x$ = 950 nm (in the propulsion direction) corresponding to a deflection angle $\theta \approx 57°$. Combined numerical and experimental optimization of metasurface parameters (see **Methods**), to simultaneously ensure maximal lateral momentum transfer and in-plane torque, resulted in dimer antenna lengths $L_1$ = 400 nm, $L_2$ = 270 nm; width $W$ = 200 nm; height $h$ = 460 nm; gap distance $g$ = 50 nm (**Supporting Figure S2**). An optimized structure is displayed in the insets of **Figure 1**. Below we focus on results for metavehicles with overall dimensions 12x10x1 µm³, though other sizes were also investigated.

**Figure 2** displays optical properties crucial to propulsion and steering for the case of an optimized metavehicle. For a normal incidence near-infrared plane wave polarized parallel to the nanoantenna's long axis, calculations indicate that the metavehicle is able to deflect ~60% of the intensity into the transmitted +1 diffraction order (directed towards the elongated antenna element), while less then 10% (5%) of the intensity falls into the 0 ( -1) orders (**Figure 2a** left). The remaining intensity is reflected, with ~15% to the +1 diffraction order and ~5% each to orders 0 and -1 (**Supporting Figure 3a**). In contrast, transmission and reflection for $y$-polarized incidence display very small deflection anisotropy (**Supporting Figure S3b,c**). The net result of the asymmetric transmission and reflection is that the metavehicle will be driven forward only when the light's polarization is aligned with the diffractive axis of the metasurface, in which case ~70% of the total incident linear momentum along $z$ is converted to a reactive force in the positive $x$-direction (Eq. 1 in **Methods**). Experimental Fourier images (**Supporting Figure S4**), obtained at the design wavelength 1064 nm, verifies the calculated beam deflection properties for transmission (**Figure 2a** right and **Supporting Figure S3c,d**).

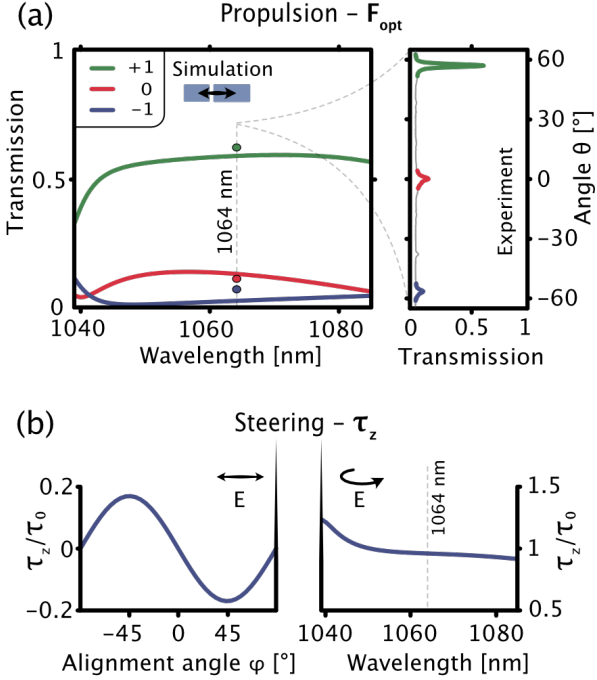

**Figure 2: Optical properties enabling metavehicle propulsion and steering. (a)** Beam-bending properties of the metasurface resulting in metavehicle propulsion. Left: Calculated near-infrared transmission into diffraction orders 0 and ±1 for normally incident light polarized along the long axis of the nanoantenna. Right: corresponding experimentally recorded diffraction efficiency measured at $\lambda_0 = 1064$ nm. **(b)** Calculated torques normalized to the torque generated by a completely absorbed circularly polarized beam ($\tau_0$). Left: Transferred torque for normally incident 1064 nm light with linear polarization aligned at angle $\varphi$ with respect to the propulsion direction. The data implies that the metavehicle experiences a restoring torque that aligns it to the polarization direction. The maximum restoring torque occurs for $\varphi = \pm 45°$. Right: The metavehicle experiences a constant torque that is independent of metasurface orientation for circularly polarized light.

We calculated the *z*-component of the integrated total spin density and extracted the reaction torque to verify that the optimized geometry also provides the expected torque in response to polarization (**Methods**). For linearly polarized light, the spin angular momentum transfer depends on the alignment angle $\varphi$ between the nanoantenna's long axis and the polarization direction, and results in a restoring torque that strives to align the metavehicle along the input polarization (**Figure 2b** left). This implies that the metavehicle will self-correct its orientation to maximize the reactive force due to linear momentum transfer. The effect is maximized at $\varphi = \pm 45°$, as for a dipole in an applied static field. For circularly polarized incidence (**Figure 2b** right), the metavehicle experiences a torque in the same direction as the handedness of the incoming light. For symmetry reasons, this torque is independent of alignment. However, since the circularly polarized wave will also be partly deflected, the combined angular and linear momentum transfer will result in orbital motion (i.e., the net effect is a spin-to-orbital angular momentum transfer).

Translation and rotation dynamics were examined with metavehicles sedimented on the bottom glass of a thin sample cell filled with water. A 1064 nm laser beam loosely focused to a diameter of ~0.4 mm and incident from above emulate a plane wave (**Methods** and **Supporting Figure S5**). The metavehicles can travel freely along the glass/water interface without the need to move the microscope stage nor the illumination position. Optical gradient forces and thermal effects are negligible for the power levels used (light intensity $I < 20\ \mu W/\mu m^2$, or power per metavehicle $P < 2\ \mathrm{mW}$).

As anticipated from the design considerations, the metavehicles propagate along straight lines and move a distance proportional to the time-integrated applied intensity when the incident laser polarization is linear (**Figure 3a**). For circular polarization, the metavehicles instead continuously turn while being propelled, with the orbital direction determined by the handedness of the incident light (**Figure 3b,c**). Tracking of metavehicle movement for increasing laser power showed that the average speed increases linearly with light intensity (bottom panels of **Figure 3d,e**), but with a ~30% lower proportionality constant for orbital motion due to time-averaging between the polarization which drives (along the metasurface's $x$-axis) and the orthogonal non-driving one. We found no noticeable difference in propulsion or steering ability between metavehicles facing up or down on the substrate. However, there are fluctuations in speed along a given trajectory, likely due to local variations in particle-surface interactions (see **Supporting Figure S6** for typical particle time-traces). The orbital radius similarly varies between metavehicles (top panel in **Figure 3e**), but the average is found to be independent of applied intensity since the linear and angular momentum transfer are both proportional to the incident intensity (top panel in **Figure 3e**).

Furthermore, we quantified the metavehicle's ability to align to a linear polarization by measuring the alignment angle $\varphi$ along various trajectories. The average $\varphi$ is found to vary within the same small interval (~±8°) irrespective of applied intensity (top panel in **Figure 3d**). As discussed above, this implies that propulsion is self-rectifying

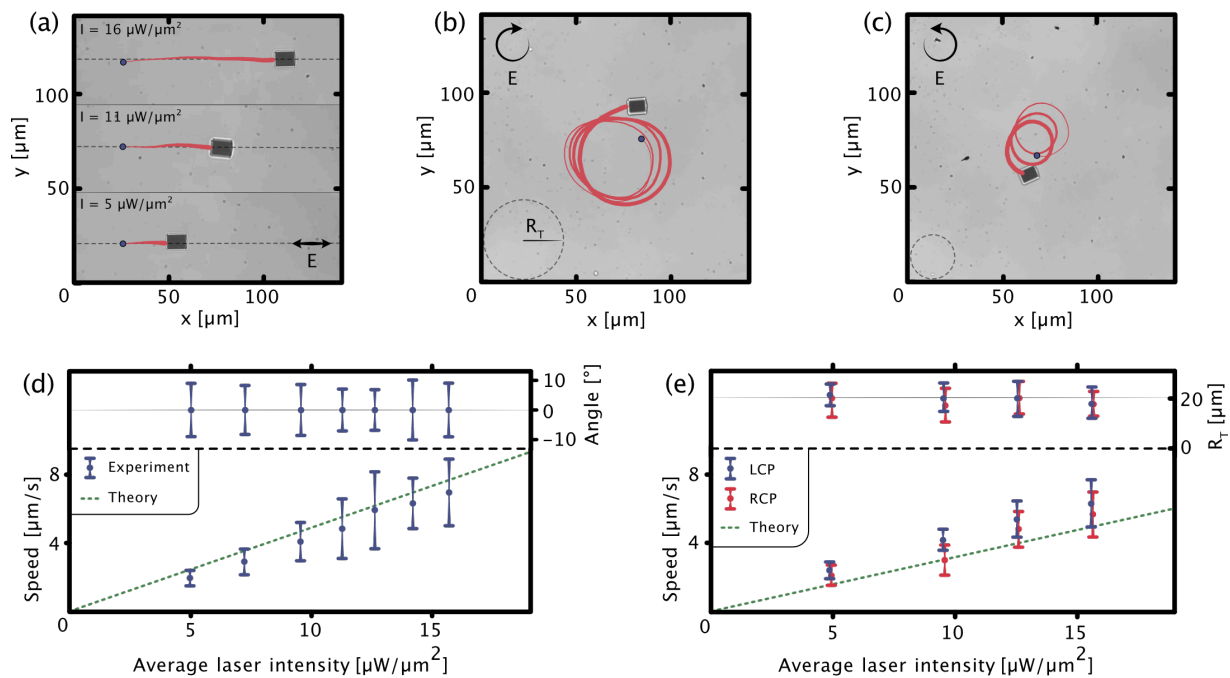

**Figure 3: Basic driving modes. (a)** The metavehicles move along straight lines parallel to the polarization direction of an incident linearly polarized laser field and travel a distance proportional to the applied time-integrated intensity (each panel shows 15 sec. of travel for one metavehicle). **(b,c)** The metavehicles travel in orbits for circularly polarized incidence with the sense of rotation determined by the polarization handedness. The dashed grey circles in the lower left corners indicate the average turning radii for each metavehicle trace. **(d, e)** Bottom panels: mean speed and standard deviation averaged over seven metavehicles versus laser intensity for **(d)** linearly polarized and **(e)** left-handed and right-handed circularly polarized (LCP, RCP) illumination. The top panel in **(d)** shows the standard distribution around the mean alignment angle $\varphi$ with respect to the forward travelling direction for linear polarization, while the top panel in **(e)** shows the mean turning radius of the metavehicles. The dashed lines in **(d, e)** show a theoretical estimate of the speed based on the force balance between the calculated optical driving force and the measured hydrodynamic drag.

against small disturbances that would otherwise cause the metavehicle to deviate from the intended direction of travel. Additional control experiments verified that metavehicles containing symmetric dimer antennas remain stationary (as a consequence of the absent asymmetric diffraction) while exhibiting polarization alignment for linearly polarized incidence and continuous spinning around their own axis for circularly polarized light (**Supporting Figure S7**).
6

The speed acquired by a metavehicle in water at low Reynolds numbers is set by the balance between the optical propulsion force and the viscous drag force, yielding $v = \frac{\partial P_x}{\partial t}/C_D$ with $P_x$ being the linear momentum transfer from Eq. 1 in **Methods** and $C_D$ the drag coefficient. We estimated the average drag coefficient to $C_D \approx 4.7 \cdot 10^{-7}$ kg/s by a simple experiment in which the whole sample cell was tilted and the metavehicles steady-state sliding speed along the surface was measured (**Supplementary Figure S8a**). Using this value together with calculated values of optical linear momentum transfer versus intensity yields the dashed lines in **Figure 3d,e**, which agrees well with the measured metavehicle speeds for both linear and circular polarization. Data for metavehicles with ~4 times larger area also agrees well with theory (**Supplementary Figure S9**). Finite-element simulations of the flow around a moving metavehicle indicate that the experimental $C_D$ value corresponds to a physically reasonable surface separation distance of around 320 nm (**Methods** and **Supplementary Figure S8b**).

The metavehicles can be made to drive in complex patterns by changing the polarization state of the incident light during propagation. **Figure 4a,b** and **Supporting Videos S1,2** gives examples of two trajectories traced out by manually shifting between linear and circular polarization at discrete points in time during the movement. Automated computer-driven feedback and more sophisticated intensity and polarization control would allow even more advanced navigation. Moreover, since the illuminated area is much larger than a single metavehicle, it is possible to drive several vehicles in parallel (**Figure 4c** and **Supporting Video S3**). Further increasing the concentration of metavehicles in the suspension would enable large numbers of particles to simultaneously travel over a surface, all driven by the same incident light field, providing an interesting platform to study inter-particle interactions and active matter.

We finally examined the metavehicles ability to transport other types of microscopic objects introduced into the suspension. **Figure 4d** and **Supporting Video S4** demonstrate controlled lateral pushing of a ~4.2 μm diameter polystyrene (PS) bead over a distance of ~0.2 mm. The surrounding beads here serve as tracer particles to attest that no other significant forces or flows are present in the environment. This type of transport event usually ends with the bead sliding off the front of the metavehicle, a problem that is likely to be avoided by designing the front face to fit the shape of the intended cargo. **Figure 4e** and **Supporting Video S5** gives a second example of transport, here involving a single baker's yeast cell (*Saccharomyces cerevisiae*). Because of their low relative density, the cells tend to slip over the thin metavehicles when mixed in pure water. However, this could be countered by performing experiments in a weak salt solution (40 mM NaCl) to slightly decrease the electrostatic repulsion between the cells and the substrate surface. We note that the use of NIR illumination and metavehicles exhibiting negligible optical absorption greatly reduce the risk of biologically harmful photodamage or heating of the cells.

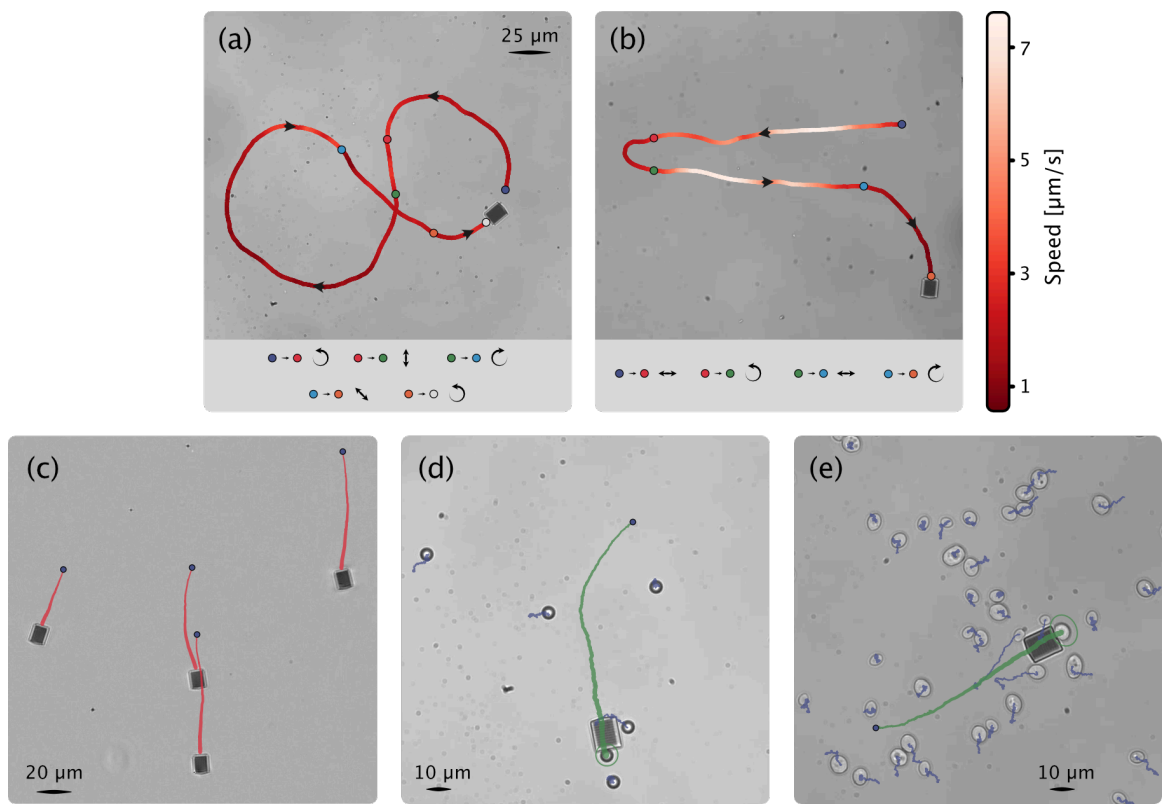

**Figure 4: Metavehicles in complex and parallel motion and as transport vehicles. (a,b)** The metavehicles can be guided to follow user-controlled trajectories by controlling the polarization state of the incident light during travel. The line color of each trace indicates the instantaneous speed, while the lower panels indicate the applied polarization state. The averaged laser intensity in the beam was 7.5 µW/µm² in (a) and 16 µW/µm² in (b). **(c)** Example of several metavehicles being driven in parallel at an intensity of 16 µW/µm² and linear polarization oriented in the vertical direction of the image. **(d,e)** The metavehicles can function as transport vehicles for cargo delivery on an interface. The objects being transported, marked by green rings, are a ~4.2 µm diameter polystyrene microbead in (d) and a yeast cell in (e). The cargo trajectories are shown in green while position fluctuations of adjacent objects, due to Brownian motion, are indicated by blue traces. The applied intensity was 16 µW/µm² in both (d) and (e).

Both the PS beads and the yeast cells slightly increase the overall drag coefficient of the metavehicle-object system, but the effect is too small to cause a noticeable slowdown of movement. The metavehicles can also transport much larger objects, as illustrated in **Supporting Video S6** where a dust particle about 15 times the size of a metavehicle is pushed over a considerable distance, albeit with somewhat impaired steering efficiency.

In summary, we have shown that it is possible to construct metavehicles that are able to propagate freely across a surface in plane-wave illumination and that can be steered by control of the incident polarization. The basic idea is related to the recently proposed theoretical concept of self-stabilizing photonic levitation of nanostructured macroscopic objects[11] in the sense that both rely on an ability to self-correct the orientation, and hence the motion, using optical torques. Artificial micro- and nanostructures designed to convert optical linear and angular momentum into mechanical motion have previously been realized in several formats, including stationary optically driven micromachines,[12,13] rotary nano- and micromotors that spin in focused light fields,[10,14,15] and plasmonic objects propelled by anisotropic scattering in line foci[16,17]. However, except for a few studies in which microscopic objects redirect plane waves without steering capabilities[18–20], all previous studies utilized gradient forces from laser beams focused to a point or a line to confine movement to a specific track or position.

Going forward, the multitude of possible optical metasurfaces offers a wealth of opportunities to introduce novel functionalities to light-driven objects. For example, Pancharatnam-Berry phase-gradient metasurfaces[2,21] could be used to control propagation through the handedness of polarization; metavehicles constructed as high-numerical-aperture flat lenses could function as miniature laser tweezers;[4] and spatially multiplexed[22] or colour-routing[23] metasurfaces could be used to build metavehicles capable of altering their functionalities depending on illumination wavelength. Metavehicles could also include molecular sensing functionalities[24] or allow for additional complementary propulsion mechanisms.[15]

# METHODS:

### Theoretical basis for beam-bending and spin transfer

Consider a two-dimensional periodic structure with lattice constants $\Lambda_x$ and $\Lambda_y$ embedded in a homogeneous isotropic host medium with refractive index $n_b$. According to Bloch's theorem, a normally incident plane wave is scattered into a number of diffraction orders that can be labelled by the in-plane wave vector $\mathbf{k}_{\parallel} = \left(\frac{2\pi n}{\Lambda_x}, \frac{2\pi m}{\Lambda_y}\right)$. For incident light with free space wavelength $\lambda$, only orders with $|\mathbf{k}_{\parallel}| < \frac{2\pi n_b}{\lambda}$ are propagating. To enable propulsion along the $x$-direction while preventing it along the $y$-direction, the metasurface lattice is designed to have $\Lambda_y < \frac{\lambda}{n_b} < \Lambda_x$. This allows propagating diffraction orders in the $m = 0$ subspace while suppressing all $m \neq 0$ orders.

Let $\mathbf{e}_{n,m}$ be the complex electric field amplitude of the transverse wave in propagating diffraction order $(n, m)$. The in-plane linear momentum density carried by each diffraction order within the $m = 0$ subspace is $p_x = \frac{1}{2Zc}|\mathbf{e}_{n,0}|^2 \sin\theta_{n,0}$, where $Z = \sqrt{\mu_0/(\varepsilon_0 n_b^2)}$ is the impedance of the host medium and $\theta_{nm}$ is the angle between the normal and the propagation direction (SI units are used). Taking into account the surface area $A$ of the object and collimation (narrowing) of the diffracted beam cross-section by a factor $\cos\theta_{n,0}$ (a beam propagating near grazing angle from a finite area grating has a vanishing cross-section), the total in-plane momentum carried by each order becomes $P_x = \frac{1}{2Zc}|\mathbf{e}_{n,0}|^2 A \sin\theta_{n,0} \cos\theta_{n,0}$. Therefore, by designing the metasurface such that $\mathbf{e}_{n,0} \neq \mathbf{e}_{-n,0}$, we can enable the in-plane momentum transfer from the electromagnetic field to the object.

To enable steering of the metavehicle by incident polarization, the metasurface is designed to be anisotropic, i.e. having different responses for $x$- and $y$-polarized incident light. This anisotropy allows converting the spin of the incident light, thus inducing torque (spin-transfer) on the structure. Our aim is that the structure should align with the incident polarization state of light that causes the in-plane propulsion. To simultaneously fulfil the two requirements of propulsion and steering, we consider a highly elongated geometry for the metasurface unit cell. Next, we optimize the unit cell to exhibit a maximal diffraction asymmetry and at the same time a minimal asymmetry for $y$-polarized light, thus preventing drift in the unwanted direction.

The unit cell of the metasurface consists of two rectangular poly-Si blocks with dimensions $L_i \times W \times H, i = 1,2$ separated by a face-to-face gap $g$ (**Supplementary Figure S2a**). The asymmetric diffraction is achieved by breaking the mirror symmetry of the unit cell with respect to the $yz$ vertical plane by requiring $L_1 \neq L_2$.

To assess the in-plane momentum carried by the $m = 0$ diffracted orders, we numerically calculate transmitted and reflected intensities $T_n$ and $R_n$ of each propagating diffraction order (assuming that the momentum carried by the evanescent orders is much smaller because of their finite transverse extent). Recalling that transmission coefficients $T_n$ describe projection of the time-averaged Poynting vector of scattered orders $\langle \mathbf{S}_n \rangle$ on the z axis, $T_n = \langle S_n \rangle \cos \theta_{n,0} / \langle S_{inc} \rangle$, and combining with the Poynting vector of each diffraction order $S_n = \frac{1}{2Z} |\mathbf{e}_{n,0}|^2$, we obtain an expression for the total in-plane momentum transfer per unit area of the object:

$$\frac{P_x}{A} = \frac{\langle S_{inc} \rangle}{c} \sum_n (T_n + R_n) \sin \theta_{n,0}. \tag{1}$$

Naturally, the specular term with $n = 0$ vanishes from this series thanks to $\sin \theta_{0,0} = 0$.

The spin transfer from a linearly polarized normally incident wave is assessed by calculating the time-averaged spin angular momentum (SAM) densities $\mathbf{s}_{t,r}$ of the total transmitted and reflected fields:[25]

$$\mathbf{s}_{t,r} = \frac{1}{4\omega} \text{Im}[\varepsilon \varepsilon_0 \mathbf{E}^* \times \mathbf{E} + \mu \mu_0 \mathbf{H}^* \times \mathbf{H}]$$

Spin continuity[26] allows us to calculate the total scattered SAM per unit area by integrating the SAM density over the unit cell $\Lambda$ in a horizontal plane away from the structure, and adding up the transmitted and reflected contributions:

$$\frac{\langle s_{scat} \rangle_z}{A} = \frac{1}{\Lambda_x \Lambda_y} \int_\Lambda \langle s_t + s_r \rangle_z \mathrm{d}x \mathrm{d}y \tag{2}$$

When the structure is excited by linearly polarized light, the incident SAM is zero, and the spin transferred to the structure is minus the total scattered spin, whereas for circularly polarized light is simply the difference between the incoming and outgoing SAM. Thus, the torque $\tau_z$ exerted on the structure along the vertical axis per unit area is the SAM flux:[27]

$$\frac{\tau_z}{A} = -c \frac{\langle s_{scat} \rangle_z}{A}.$$

### Simulations

Electrodynamics simulations of the metasurface optical properties were performed with a finite-difference time-domain (FDTD) method using commercial software (Lumerical, Canada). Transmission and reflection spectra of diffraction orders and SAM densities were obtained using a normally incident linearly polarized plane-wave source and in-plane periodic boundary conditions. Polycrystalline silicon was modelled with a constant real-valued refractive index $n_{Si} = 3.45$ since absorption is negligible at the design wavelength ($\text{Im}\{\varepsilon_{\text{Si}}\} < 0.003$). Maximization of diffraction asymmetry was done with the use of the built-in particle swarm optimization algorithm using 50 generations and a population of 20 for each generation. The free parameters for this optimization was the antenna lengths and width, and non-diffractive unit cell size, whereas the remaining parameters were fixed by fabrication constrains.

Computational fluid dynamics simulations to estimate the drag coefficient of a metavehicle subject to translation near a glass-water interface were performed using COMSOL Multiphysics. The simulation environment utilized the laminar flow approximation and consisted of a 200x200x200 µm³ bounding region. An inlet and outlet were placed on the exterior $yz$ planes with flow velocity $\vec{u}_0$ = 10 µm/s directed along the $x$-axis, whereas all other boundaries were set as sliding walls with velocity $\vec{u}_0$ to establish a uniform background fluid flow profile in the metavehicle's frame of reference. The metavehicle was created in the center of the $xy$ plane with variable separation distance from the lower $xy$ bounding plane. All surfaces of the metavehicle were specified as no-slip boundaries. A local high-density mesh region extending twice the thickness from the metavehicle was used to ensure accurate results in the high velocity gradient region, especially for small separation distances. The stress parallel to $\vec{u}_0$ was integrated over the surface of the metavehicle to determine the total effective drag force, which was then divided by the normal inflow velocity to obtain the drag coefficient.

Fabrication

The metavehicles were fabricated using a top-down approach with dual exposure electron beam lithography (EBL) and subsequent etching of a sacrificial substrate. As graphically illustrated in **Supporting Figure S1,** the fabrication was initiated by depositing amorphous silicon (a-Si) by low-pressure chemical vapor deposition on 400 nm thick thermally grown $SiO_2$ on a 4-inch Si wafer. The a-Si was then annealed for 30 minutes at 1000°C in an inert atmosphere, which resulted in a poly-crystalline Si film of 460 nm thickness. An EBL step was performed to define the metasurface features in a negative resist mask (ma-N 2403 baked at 90°C for 90 s, exposed at 580 µC/cm2 with a current of 100 nA, and developed for 6 min). The pattern could then be transferred to the poly-Si by reactive ion etching with HBr (80 sccm HBr, 2 mTorr, 500/250 W ICP/FW). The remaining resist mask was removed by oxygen plasma stripping (10 min at 10 sccm O2, 200 mTorr, 250 W). The poly-Si metasurface was then covered by a 600 nm $SiO_2$ layer deposited using plasma-enhanced chemical vapor deposition (400 sccm 2% $SiH_4$ and 1420 sccm $N_2O$, 500 mTorr, 300°C). A second EBL step was performed to expose the metavehicle shape in a positive resist mask (ARP6200.13 spun at 1000 rpm and baked at 160°C for 5 min, exposed at 1200 µC/cm2 with a current of 100 nA, and developed for 12 min), which was then used to lift off the superfluous Ni deposited as a mask for the subsequent etch step. Next, the metavehicles were excavated using reactive ion etching, where both the exposed PECVD and thermal $SiO_2$ were removed (10 sccm $CHF_3$ and 15 sccm Ar, 5mTorr 600/20 W ICP/FW). This leaves the metavehicles attached to the Si wafer. To release these into a colloidal solution, the high selectivity between Si and $SiO_2$ for reactive ion etching with $SF_6$ was utilized to fully etch the attached Si (15 min at 50 sccm $SF_6$, 40mTorr, 300/10 W ICP/FW), and hence leave the metavehicles scattered on the substrate. The fabrication was finalized by sonicating the sample in de-ionized water to release the metavehicles from the substrate and disperse them in solution.

Experimental optimization of metasurface optical properties was performed using samples produced on transparent 4-inch fused silica wafers but not released into solution (fabrication steps 1-3 in **Supporting Figure S1**). **Supporting Figure S2** summarizes experimental variation of critical metaatom dimensions ($L_1$, $L_2$, W, g).

**Optical measurements**

The optical characterization setup is illustrated in **Supporting Figure S4.** Metasurfaces were illuminated with linearly polarized, quasi-monochromatic and collimated 1064 nm light at normal incidence. The transmitted light was collected using a water immersion objective with high numerical aperture and then projected on a Fourier plane to allow for quantification of polarization-dependent diffraction and overall transmission efficiency.

Optical actuation of metavehicles in water was achieved using 1064 nm laser illumination. The laser beam had a spot size diameter of ~400 μm (FWHM) and a divergence angle ~0.025 deg. at the sample, thus approximating a plane wave within the microscope field of view. A quarter-wave plate and half-wave plate allowed the polarization state to be changed between linearly polarized at an arbitrary direction or circularly polarized with either right or left handedness. Measurements were performed using a ~4 μl sample cell formed by two microscope slides separated by a 110 μm spacer. The microscope slides were made hydrophilic by cleaning in 2 wt% Hellmanex III heated to 70°C for 15 minutes to prevent the metavehicles from sticking to the glass substrate. Video tracking analysis allowed for quantification of metavehicle position and orientation changes.

## CONFLICT OF INTEREST:

The authors claim no conflict of interest.

## AUTHOR CONTRIBUTIONS:

M.K. & D.A. conceived the study. D.A. manufactured samples, performed experiments, and analyzed data. D.G.B. performed electrodynamics simulations and optimization. S.J. performed hydrodynamic simulations. D.A., M.K. and D.G.B wrote the paper with input from all authors.

ACKNOWLEDGEMENTS:

This work was supported by the Knut and Alice Wallenberg Foundation, the Swedish Research Council and the Excellence Initiative Nano at Chalmers University of Technology. The nanofabrication for this work was performed at Myfab Chalmers.

SUPPORTING INFORMATION:

Supporting information is available free of charge.

# Supporting Information:

# Microscopic Metavehicles Powered and Steered by Embedded Optical Metasurfaces

Daniel Andrén[†,*], Denis G. Baranov[†], Steven Jones[†], Giovanni Volpe[‡], Ruggero Verre[†], Mikael Käll[†,*]

[†]*Department of Physics, Chalmers University of Technology, S-412 96 Gothenburg, Sweden*

[‡]Physics Department, Gothenburg University, *S-412 96 Gothenburg, Sweden*

*Corresponding authors: daniel.andren@chalmers.se, mikael.kall@chalmers.se*

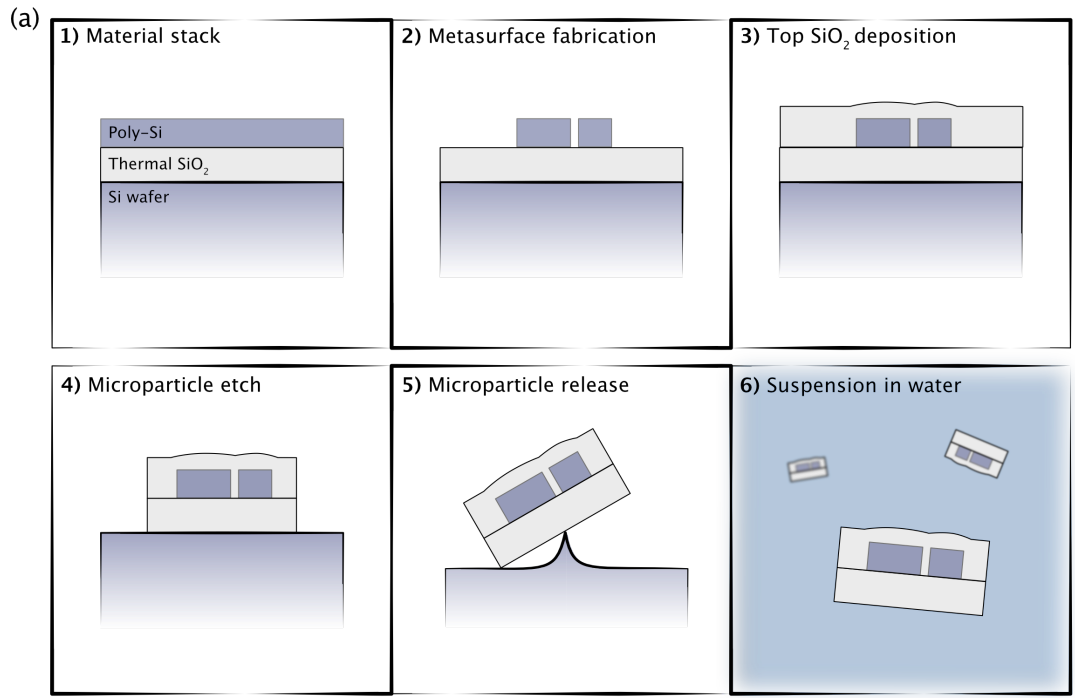

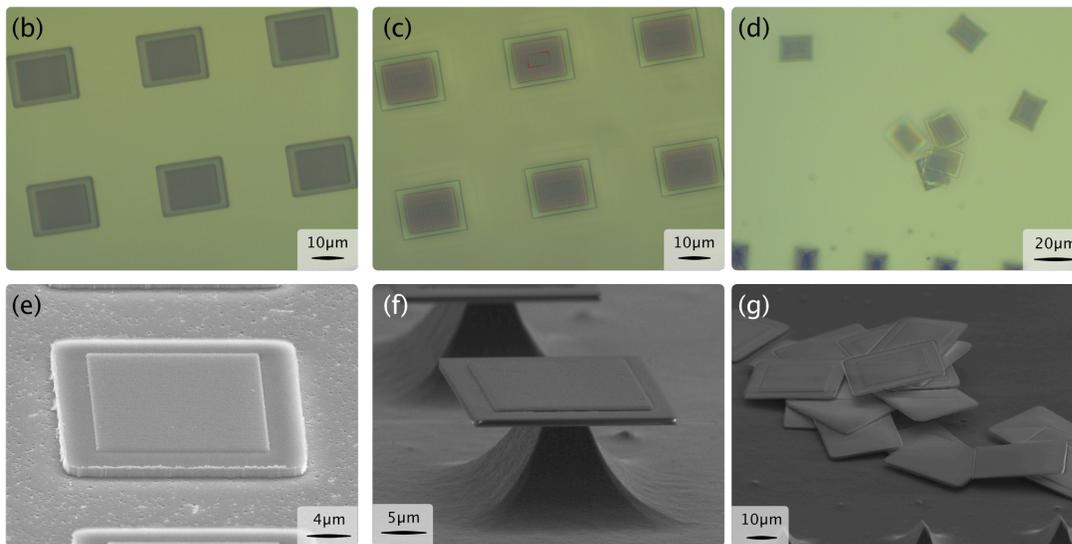

**Figure S1: Fabrication of metavehicles. (a)** Illustration of critical fabrication steps: 1) Initial material stack, consisting of a 4-inch Si wafer, 400 nm thermally grown $SiO_2$, and 460 nm poly- Si; 2) Definition of metasurface features in a negative resist (ma-N 2403) mask using electron-beam lithography (EBL, Raith EBPG 5200), followed by pattern transfer to poly-Si through reactive ion etching in HBr; 3) Encapsulation of the poly-Si metasurface by a 600 nm thick conformal $SiO_2$ layer; 4) Excavation of the structure in a second EBL step followed by reactive ion etching; 5) Release of the metavehicles from the Si substrate using reactive ion etching with $SF_6$; 6) Suspension of the metavehicles in water. **(b)** Optical micrographs of metavehicles before Step 4, **(c)** before Step 5 (remaining Si pillars, of which one is highlighted by a red rectangle, can be seen through the metasurfaces), and **(d)** before Step 6. **(d-f)** Scanning electron micrographs acquired during the same fabrication stages as shown in (b-d).

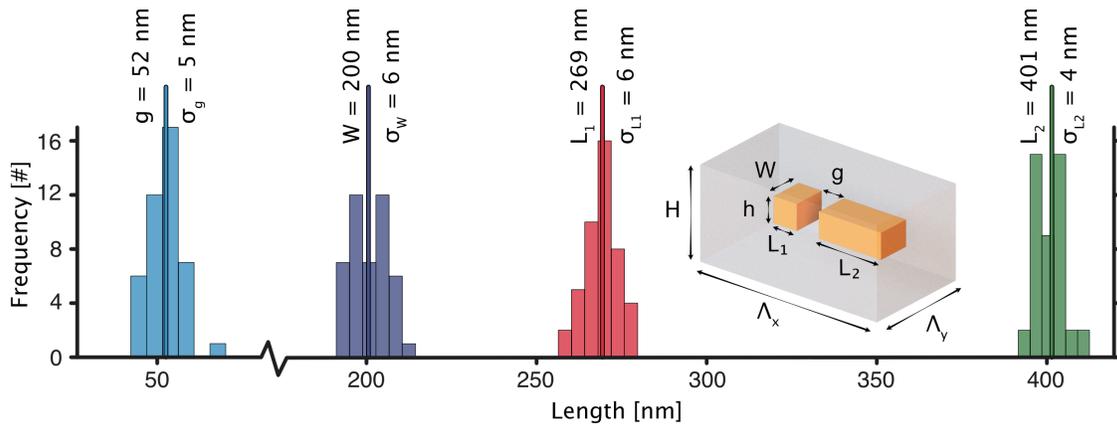

Figure S2: Critical metasurface dimensions. (a) Histograms of measured directional nanoantenna dimensions critical to metavehicle functionality, together with corresponding mean values and standard deviations.

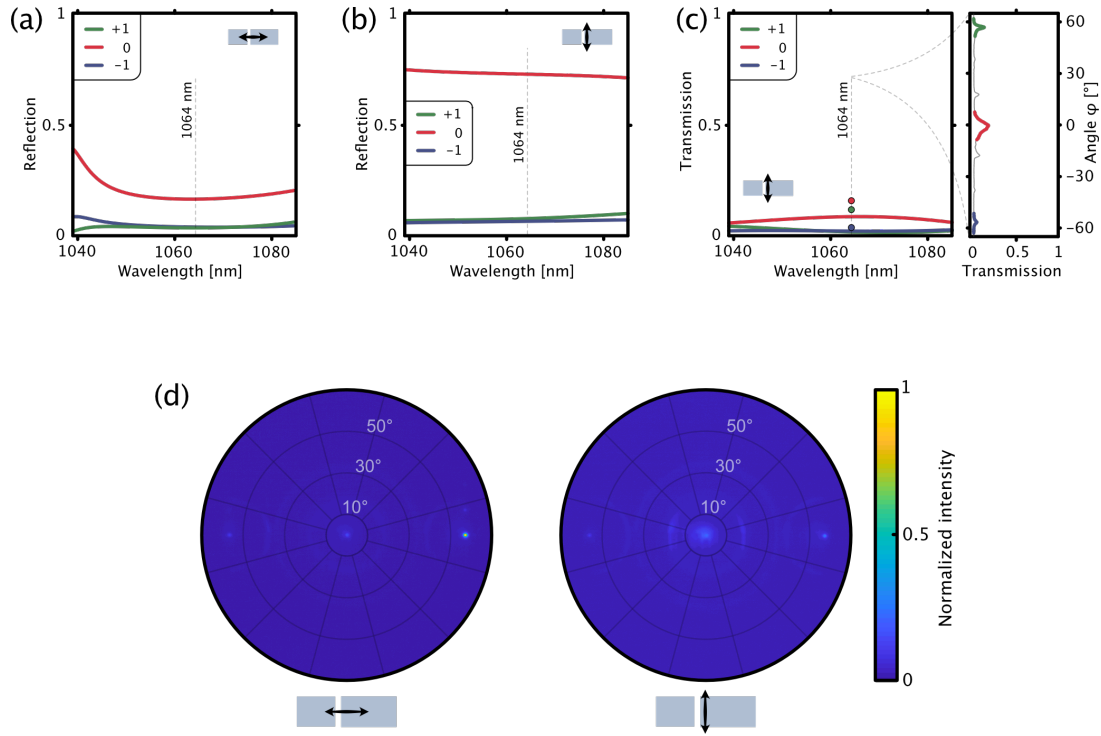

Figure S3: Optical properties enabling metavehicle propulsion and steering. (a) and (b) present reflection spectra for n = 0th and ±1st diffraction orders, for light polarized along the long and short axis of the meta-atoms, respectively. (c) Same data as presented in main **Figure 2a**, but for polarization along the short axis of the meta-atom. That is, simulated transmission spectra for the n = 0th and ±1st diffraction orders, as well as line-averaged experimental data extracted from the right panel of of (d). (d) Experimental Fourier images of the light transmitted through the metasurface at 1064 nm for polarization along the two symmetry axes. For light polarized along the long axis of the meta-atom (left) a clear transmission asymmetry is observed (this image is the source of the experimental data presented in **Figure 2a**), whereas light polarized along the short axis of the meta-atom (right) is close to symmetric. Both graphs share the same color scale, which is normalized to the maximum intensity. Note that due to the sub-diffraction limited unit-cell of the metasurface in the $y$-direction ($\Lambda_y < \lambda/n_b$) no propagating diffraction orders are observed in the y-direction. The faint light present here is attributed to diffuse scattering from defects in the metasurface sample.

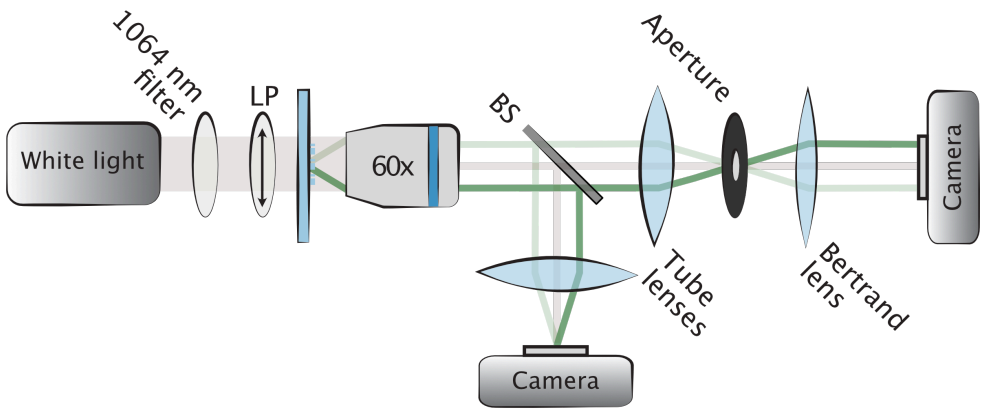

**Figure S4: Optical setup for measuring metasurface directionality.** White light from a laser-driven light source (Energetiq EQ- 99XFC LDLS) is collimated and filtered via a narrow bandpass filter centered at 1064 nm (Thorlabs FL1064-10), and then linearly polarized (LP), before hitting the sample at normal incidence. The transmission is collected with a 60x NA 1.2 water immersion objective (Nikon CFI Plan Apo VC 60XC WI), then split in two parts using a non-polarizing beam splitter (BS). Images of the sample are projected on a camera, for orientation, and on an aperture, used to isolate the metasurface transmission from the background light not passing through the metasurface. A Bertrand lens is used to form a Fourier image of the transmission on a second camera (Thorlabs DCC1545M).

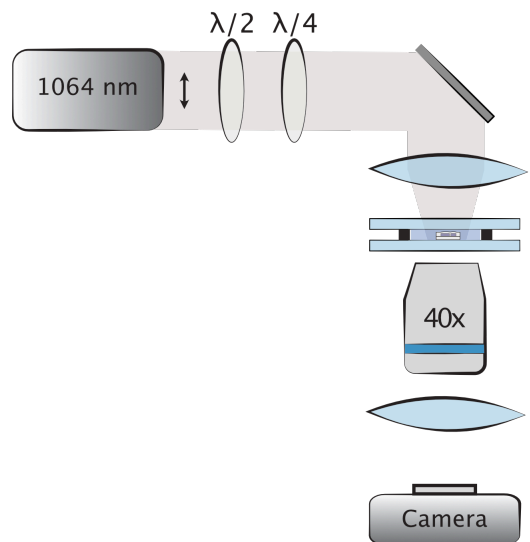

**Figure S5: Optical setup for driving and observing metavehicle motion.** A 1064 nm CW laser (Cobalt Rumba 2W) beam is expanded to a diameter of ~1 cm and focused through a 20 cm lens to a spot size of ~0.4 mm. A combination of half- and quarter-wave plates allows the polarization state to be changed between linearly polarized at any arbitrary direction or circularly polarized with either right or left handedness. The metavehicles are contained in a thin liquid sample cell, where they sediment on the lower glass surface, and illuminated from above. The metavehicle movement is imaged by a 40x objective (Nikon CFI Plan Apo Lambda 40X NA 0.95) and a CMOS camera.

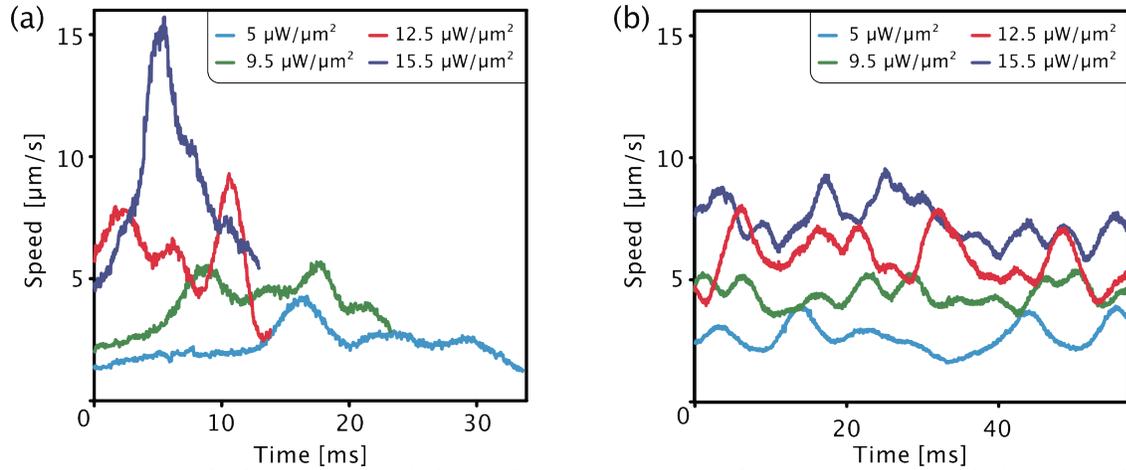

**Figure S6: Time-resolved motion.** Metavehicle speed versus time extracted from tracking two individual motors as they are propelled at different laser intensities of **(a)** linearly and **(b)** circularly polarized light. The metavehicle eventually exits the observation area in the case of linear polarization, resulting in a shorter observation time for higher laser powers (higher average speed). In the case circularly polarized illumination, the particle orbits within in the illuminated area, meaning that tracking can be performed for extended periods. The velocity fluctuations are attributed to inhomogeneities in the metavehicle – substrate interaction, primarily due to variations in hydrophilicity and surface charge (the metavehicles stick to substates that have not been washed in Hellmanex, but surface variations may prevail even after washing).

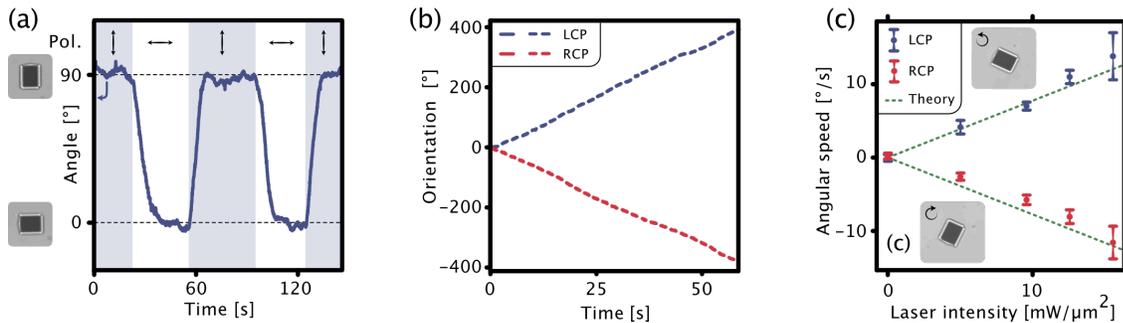

**Figure S7: Control experiments on symmetrized metavehicles. (a)** Orientation versus time for a metavehicle containing symmetric dimer antenna elements (with the same Si volume as the directional antennas and with otherwise identical structure parameters) at an incident intensity of 16 µW/µm². The metavehicle orientation follows the direction of the alternating linear polarization but there is no directional translation. **(b)** For circularly polarized light, the symmetrized metavehicle remains at the same position while continuously spinning around its own axis in a direction dictated by the handedness of the incident light (LCP/RCP stands for left-/right-handed circular polarization). **(c)** The angular speed increases linearly with the applied intensity for circularly polarized incidence. The optical torque (main **Figure 2b**) is balanced by a rotational drag force $(F_{D,R} = C_R \Omega)$ to produce an angular velocity of $\Omega = \frac{\tau_z}{C_R} = \frac{IA\lambda}{2\pi c \cdot C_R}$. A value for the rotational drag coefficient of ~5·10⁻⁷ Js produces a theoretical trend (dashed green line) that agrees well with the measured rotation speeds. Assuming that the asymmetric metamotors experience a comparable torque, this angular speed together with the linear speed ($v$) of the driven metamotors under circular polarization (main **Figure 3e**) allow us to estimate their turning radius as $R_T = \frac{v}{|\Omega|}$. This is found to be around 25 µm and hence consistent with the findings presented in the top panel of **Figure 3e**

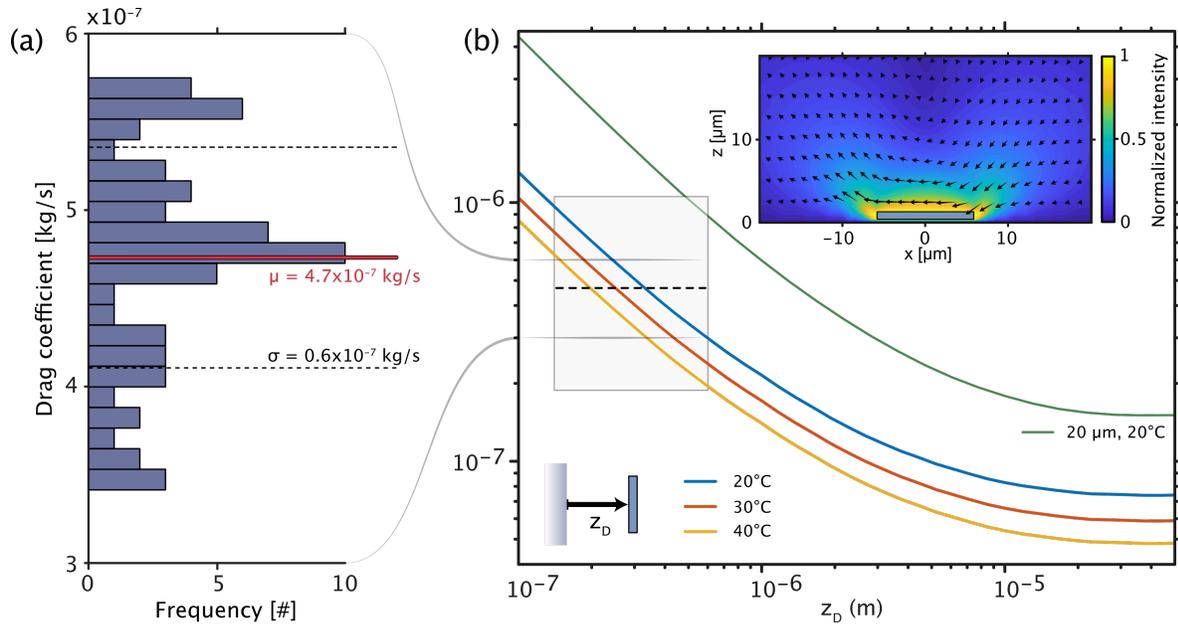

Figure S8: Hydrodynamic interactions at the glass/water interface. (a) Experimental estimations of the drag coefficient for metavehicles sliding along the surface of a tilted sample cell due to gravity (without any incident laser field). The mean value and standard deviation of the drag coefficient distribution are indicated by the red and dashed lines respectively. (b) Finite element simulations (COMSOL Multiphysics) of the drag experienced by a cuboid object with the same dimensions as the metavehicles (1x10x12 µm³) as it translates in water along a glass interface. The simulation is performed at increasing surface-to-surface separation distance $z_D$ and at three different bath temperatures. The dashed line denotes the mean value of the experimentally extracted drag from (a) whereas the grey lines denote the range of the histogram. The grey box indicates what is interpreted as physically reasonable separation distances and drag values. The solid green line represents the drag experienced by a larger metavehicle (dimensions 1x19x21 µm³). The inset displays the simulated velocity flow profile around a metavehicle at $z_D$ =316 nm.

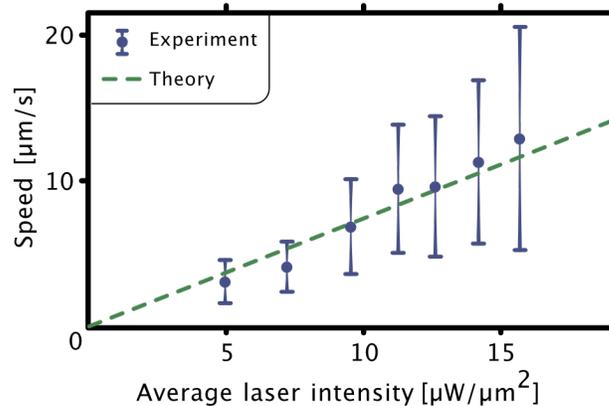

Figure S9: Speed versus laser intensity for a larger metavehicle. Experimental data for metavehicles with size 21x19x1 µm³ together with calculated speed versus intensity based on the drag coefficient shown in **Figure S8b** (assuming a separation distance of ~350 nm). Due to the four times larger metasurface area, these metavehicles are propelled by an approximately four times larger optical force than the smaller metavehicles investigated in the main text. However, this occurs with a concomitant increased drag, ultimately resulting in a speed that is about twice that of the smaller case for the same applied intensity.